\documentstyle[a4,11pt]{article}

\begin{document}
\begin{center}
{\LARGE Estimate of the Hausdorff Dimension of a Self-Similar Set due to Weak Contractions}\\~\\

{Yoshihito Ogasawara\renewcommand{\thefootnote}{*}\footnote{E-mail address: ogasawara@aoni.waseda.jp}and Shin'ichi Oishi}

{Faculty of Science and Engineering, Waseda University, Ohkubo, Shinjuku-ku, Tokyo 169-8555} 
\end{center}

{\small As for the remarkable study on the estimate of the Hausdorff dimension of a self-similar set due to weak contractions (Kitada A. et al. Chaos, Solitons \& Fractals 13 (2002) 363-366), we present a mathematically simplified form which will be more applicable to various phenomena.

}

~\\

In the field of the crystallography, a notable map $f:X\to X$, called a weak contraction, which satisfies the condition (\ref{def}) was proposed \cite{poly} where $X$ is a metric space equipped with a metric $d$ \cite{nonli}. 
\begin{eqnarray}
d(f(x),f(y))\le\alpha(t)d(x,y),~d(x,y)<t,~0\le\alpha(t)<1.
\label{def}
\end{eqnarray}
This contraction is literally "weak" in the sense that the upper limit $\sup_{t>0}\alpha(t)$ of the variable contraction coefficient $\alpha(t)$ can be 1. 
Then, owing to the variability of the coefficient, the weak contraction can describe phenomena in greater detail than the conventional one. For instance, the weak contraction $f$ can more exactly describe such an intriguing situation that there exists a nonnegative number $t_0$ such that $\alpha(t)\to 0~(t\searrow t_0)$.

Furthermore, the weak contraction $f:X\to X$ possesses one of the most important properties of conventional contractions, 
that is, $f$ has a unique fixed point if $X$ is complete \cite{poly}. By the use of this property and the set dynamics, it is exhibited \cite{poly} that there exists a unique compact set $S$ in $X$ such that $\bigcup_{j=1}^mf_j(S)=S$ if there exist $m$ weak contractions $f_j:X\to X,~j=1,\ldots,m~(2\le m<\infty)$ where $X$ is complete \cite{domain}. Then, the weak contractions can be regarded as conventional contractions on the compact set $S$. Indeed, since $S$ is bounded, the relation $d(f_j(x),f_j(y))\le\alpha_j({\rm dia}S+\delta)d(x,y)$ holds for each $j$ and any two points $x$ and $y$ in $S$, where $\delta$ is a positive number and dia$S$ denotes the diameter of $S$. Accordingly, the above unique compact set $S$ such that $\bigcup_{j=1}^mf_j(S)=S$ is a self-similar set in the conventional sense 
despite of the fact that we only assume the existence of the "weak" contractions \cite{self}. 

Then, an important theorem \cite{2002} holds for the estimate of the Hausdorff dimension \cite{Haus} of the compact self-similar set constructed by "special weak contractions", and this theorem is a starting point of the discussion of the emergence of diverse significant structures such as embedment, a decomposition space, the hierarchic structure of a dendrite, chaos, and so on \cite{2002,2003}. 

The aim of this study is to verify the following proposition which is more refined and more universal than the above theorem so that we can estimate the Hausdorff dimension of the compact self-similar set $S$ constructed by "any  weak contractions", which can lead to further applications for various phenomena. \\

{\noindent\bf Proposition.}~~The Hausdorff dimension of a compact self-similar set $S$ constructed by any weak contractions $f_j,~j=1,\ldots,m$ ($2\le m<\infty$) which have their coefficients $\alpha_j(t),~j=1,\ldots,m$ is estimated as follows \cite{estimate}:
\begin{eqnarray}
\dim_HS\le x_0,
\end{eqnarray}
where the nonnegative number $x_0$ is uniquely determined by the relation,
\begin{eqnarray}
\sum_{j=1}^m(\inf_{t>0}{\alpha}_j(t))^{x_0}=1.\label{sum}
\end{eqnarray}
Here, let $x_0$ be $0$ if $\inf_{t>0}{\alpha}_j(t)=0$ for all $j$, and let $0^x$ be $0$ for a nonnegative number $x$ otherwise.

{\noindent\bf Proof.}~~First, let us recall the definition and the fundamental property of the Hausdorff dimension \cite{HW}. The Hausdorff dimension of a subset $E$ of a metric space is defined by $\dim_HE=\sup\{p\ge 0;~H^p(E)>0\}$, where $H^p(E)=\sup_{\varepsilon>0}H^p_\varepsilon(E)$ and $H^p_\varepsilon(E)=\inf_{\{C_i\}}\sum_{i}({\rm dia} C_i)^p$. $\{C_i\}$ is a countable $\varepsilon$ cover of $E$, that is, $E\subset \bigcup_iC_i$ and  dia$C_i$ is not greater than $\varepsilon$.
Then, the relation $H^t(E)<\infty\Rightarrow\dim_HE\le t$ holds.

Next, let us verify that for each $n$, the relation $\bigcup_{j_1\cdots j_n\in W_n}S_{j_1\cdots j_n}=S$ holds where $S_{j_1\cdots j_n}=f_{j_1}\circ\cdots\circ f_{j_n}(S)$ and $W_n$ denotes the set of all words $j_1\cdots j_n$ with length $n$ on symbols $1,\ldots,m$. In fact, the relation $\bigcup_{j_1\cdots j_{n+1}\in W_{n+1}}S_{j_1\cdots j_{n+1}}=\bigcup_{i=1}^mf_i(\bigcup_{j_1\cdots j_n\in W_{n}}S_{j_1\cdots j_{n}})=\bigcup_{i=1}^mf_i(S)=S$ holds if the relation holds for $n$. Namely, for each $n$, $\{S_{j_1\cdots j_n};~j_1\cdots j_n\in W_n\}$ is a finite cover of $S$.

Then, let a function $\tilde{\alpha}_j(t)~(t\ge 0)$ be defined by $\tilde{\alpha}_j(t)=\inf_{p>t}{\alpha}_j(p)$. The function $\tilde{\alpha}_j(t)$ is obviously monotone increasing, and the relation $d(f_j(x),f_j(y))\le\tilde{\alpha}_j(d(x,y))d(x,y)$ holds. In fact, since $d(f_j(x),f_j(y))\le\alpha_j(t)d(x,y)$ for any $t>d(x,y)$, $d(f_j(x),f_j(y))\le[\inf_{t>d(x,y)}{\alpha}_j(t)]d(x,y)=\tilde{\alpha}_j(d(x,y))d(x,y)$.

Then, the relation ${\rm dia}S_{j_1\cdots j_n}\le\tilde{\alpha}_{j_1}({\rm dia}S_{j_2\cdots j_n}){\rm dia}S_{j_2\cdots j_n}$ holds. In fact, for any points $x_1$ and $x_2$ in $S_{j_1\cdots j_n}$, there exist $y_1$ and $y_2$ in $S_{j_2\cdots j_n}$ such that $x_1=f_{j_1}(y_1)$ and $x_2=f_{j_1}(y_2)$, and since  $d(x_1,x_2)=d(f_{j_1}(y_1),f_{j_1}(y_2))\le\tilde{\alpha}_{j_1}(d(y_1,y_2)) d(y_1,y_2)\le \tilde{\alpha}_{j_1}({\rm dia}S_{j_2\cdots j_n}){\rm dia}S_{j_2\cdots j_n}$, ${\rm dia}S_{j_1\cdots j_n}\le\tilde{\alpha}_{j_1}({\rm dia}S_{j_2\cdots j_n}){\rm dia}S_{j_2\cdots j_n}$. Accordingly,
$$
{\rm dia}S_{j_1\cdots j_n}\le\tilde{\alpha }_{j_1}({\rm dia}S_{j_2\cdots j_n})\cdots 
	\tilde{\alpha }_{j_n}({\rm dia}S){\rm dia}S.
$$ 
Furthermore, 
$$
{\rm dia}S_{j_1\cdots j_n}\le K^n{\rm dia}S\to 0~~(n\to\infty),
$$ 
where $K={\max_{j}\{\tilde{\alpha }_{j}({\rm dia}S)\}}$.
Therefore, for any $\varepsilon>0$, there exists $N$ such that $\{S_{j_1\cdots j_n};~j_1\cdots j_n\in W_n\}$ is a finite $\varepsilon$ cover of $S$ for any $n\ge N$ \cite{def}. 

In addition, let us define a nonnegative function $x(t)~(t\ge 0)$ by $\sum_{j=1}^m\tilde{\alpha}_j(t)^{x(t)}=1$ where let $x(t)$ be $0$ if $\tilde{\alpha}_j(t)=0$ for all $j$, and let $0^x$ be $0$ for a nonnegative number $x$ otherwise. From the monotonicity of each $\tilde{\alpha}_j(t)$, $x(t)$ is monotone increasing.

Now, let us estimate the Hausdorff dimension of the compact self-similar set $S$. If there exists a positive number $t$ such that $M(t)=\{j\in\{1,\ldots,m\};~\tilde{\alpha}_j(t)>0\}$ is empty, $S$ is a finite set, and thus ${\rm{dim}}_HS=0=x_0$. Accordingly, let us consider such a situation that $M(t)\ne\phi$ for any $t>0$ \cite{H0}. 

For an arbitrary fixed positive number $t$, there exists a positive integer $p$ such that ${\rm dia} S_{j_1\cdots j_q}<t$ for any $j_1\cdots j_q\in W_q$ and any integer $q>p$.  Then, for any $j_1\cdots j_n\in W_n$ and any $n\ge p+2$, ${\rm dia}S_{j_1\cdots j_n}$ is estimated as follows.
\begin{eqnarray*}
{\rm dia}S_{j_1\cdots j_n}
&\le &\tilde{\alpha }_{j_1}(t)\cdots \tilde{\alpha }_{j_{n-p-1}}(t)
	{\tilde{\alpha }_{j_{n-p}}({\rm dia}S_{j_{n-p+1}\cdots j_n})}
		\\&&
        \cdots {\tilde{\alpha }_{j_{n}}({\rm dia}S)}
		{\rm dia}S\\
&\le &\tilde{\alpha }_{j_1}(t)\cdots \tilde{\alpha }_{j_{n-p-1}}(t)	K^{p+1}{\rm dia}S.
\end{eqnarray*}
Accordingly,
\begin{eqnarray*}
&&
\sum_{j_1\cdots j_n\in W_n}({\rm dia}S_{j_1\cdots j_n})^{x(t)}
\\
&\le &(K^{p+1}{\rm dia}S)^{x(t)}(\sum_{j\in M(t)}\tilde{\alpha }_j(t)^{x(t)})^{n-p-1}\\
&=&(K^{p+1}{\rm dia}S)^{x(t)}(\sum_{j=1}^m\tilde{\alpha }_j(t)^{x(t)})^{n-p-1}\\
&=&(K^{p+1}{\rm dia}S)^{x(t)}.
\end{eqnarray*}
Since $(K^{p+1}{\rm dia}S)^{x(t)}$ is independent of $n$, $H^{x(t)}_{\varepsilon }(S)\le (K^{p+1}{\rm dia}S)^{x(t)}$ for any $\varepsilon>0$.  Therefore, $H^{x(t)}(S)\le (K^{p+1}{\rm dia}S)^{x(t)}<\infty$, and the estimate ${\rm{dim}}_HS\le x(t)$ is obtained.   By the arbitrariness of $t$, the relation ${\rm{dim}}_HS\le \inf_{t>0}x(t)$ holds. 

Here, each $\tilde{\alpha}_j(t)$ is continuous at $t=0$ even when $\alpha_j(t)$ is discontinuous at any $t>0$ \cite{nonli}. In fact, for any $\varepsilon>0$, there exists $t'>0$ such that $\alpha_j(t')-\tilde{\alpha}_j(0)<\varepsilon$, and for any $t''\in[0,t')$, 
$\tilde{\alpha}_j(t'')-\tilde{\alpha}_j(0)\le{\alpha}_j(t')-\tilde{\alpha}_j(0)<\varepsilon$. Therefore, $x(t)$ is also continuous at $t=0$. Consequently, from the monotonicity of $x(t)$, $\inf_{t>0}x(t)=x(0)$, and thus $x(0)=x_0$ by definition.$\Box$\\


{\noindent\large\bf Acknowledgments}

The authors are grateful to Prof. Akihiko Kitada of Waseda University and Prof. Em. Yoshisuke Ueda of Kyoto University for helpful discussions. This study was supported by the Japan Science and Technology Agency.


\begin{thebibliography}{9}
\bibitem{nonli} It must be noted that this map $f$ is permitted to have extreme nonlinearity. 
\bibitem{domain} As an application of this result, the existence of a compact domain with self-similarity is shown in a polycrystal \cite{poly}.
\bibitem{self} Namely, the existence of the weak contractions is a sufficient condition 
for the existence of a self-similar set.
\bibitem{Haus} The Hausdorff dimension is a fundamental notion in recognizing the self-similarity which is a universal property through natural phenomena \cite{mandel}.
\bibitem{estimate} Owing to the variability of the coefficients, the more exact estimate of the Hausdorff dimension is obtained.
\bibitem{def} Namely, the definition of the self-similar set guarantees the existence of its finite $\varepsilon$ cover.
\bibitem{H0} It is noted that $\inf_{t>0}{\alpha}_j(t)$ can be zero for all $j$ even if $M(t)\ne\phi$ for all $t>0$.
\end{thebibliography}
\end{document}